# A semi-empirical response function for Gamma-ray of Scintillation detector based on physical interaction mechanism

LI Zhe (李哲)[1], ZHANG Yiwen (张译文)[1], SUN Shifeng (孙世峰)[1], WANG Baoyi(王宝义)[1], WEI Long (魏龙)[1,2;1)]

[1] Key Laboratory of Nuclear Radiation and Nuclear Energy Technology, Institute of High Energy Physics, Chinese Academy of Sciences，Beijing 100049, China

[2] Beijing Engineering Research Center of Radiographic Techniques and Equipment, Beijing 100049, China

**Abstract:** Scintillation detector has lower energy resolution for Gamma-ray as compared to semiconductor detector, better spectra analysis method is essential to traditional method. A model for describing the response function of scintillation detector over the range of incident Gamma-ray energies between 0.5 and 1.5 MeV has been established and applied to fitting radiation sources spectra. Each function form for describing the feature of Gamma-ray spectra are based on the analysis of fundamental interaction mechanism. These functions are combined to form a DRF model to fit experiment spectra by weighted least squares fitting method, parameters in this model are obtained simultaneously. Gaussian standard deviation can be calculated out by an individual procedure. Validity of the DRF model is demonstrated by fitting Co-60 and Cs-137 spectra measured by CsI(Tl) detector and comparing them to the normalized equivalent measured spectrum.

**Keywords:** detector response function; scintillation detector; Gamma-ray

**PACS:** 29.30.Kv; 31.15.bu

## 1 Introduction

Useful properties of scintillation detectors are their variable decay time for various exciting particles, good plasticity, less hygroscopic and large-size crystal can be grown[1]. So, this type of crystals have been incorporated in many large solid angle detection arrays[2][3]. In such applications, energy resolution is a very essential requirement of many experiments. As compared to HPGe, Si(Li), $LaBr_3$, the energy resolution of scintillation detector is not very good[4]. Thus lead to some serious problems when take Gamma-ray spectrum analysis.

To generate spectral response function of scintillation detector is an effective method to unfolding or analysis of measured spectra. A series typical detector response function (DRF) of NaI, BGO, HPGe, Si(Li), Si(PIN) and applications are reported by Gardner, Compbell and others[5-7]. There are three methods to generate DRF; the most common and effective method is to fit analytical functions that describe the various features of measured spectra from single energy sources, obtain the best (least-squares) estimates of the parameters in the fitting functions, and then obtain (again by least-squares) these parameters as a function of source energy[8]. The spectrum of Gamma-ray has limitations restricted by the energy resolution of scintillation detector ,especially CsI(Tl) detector, and so, delicate physical effect can't be reveal obviously through the spectrum, which is the main obstacle for accurate calculation of Gamma-ray count rate or spectrum net area.

There are several advantages of using DRFs in scintillation detector measured spectrum analysis, including (1) each module of DRFs is based on the fundamental physical interaction; (2) more accurate spectrum analysis results can be obtained by fitting with DRFs. And also, DRF can

* Supported by National Key Foundation for Exploring Scientific Instrument of China(2011YQ120096); Innovative Program of The Chinese Academy of Sciences(KJCX2-EW-N06)

1) E-mail: weil@ihep.ac.cn

be made considerably more accuracy in spectral simulations and are critical in weighted least square analysis.

In this paper, we analyze the physical interaction mechanism of scintillation detector to Gamma-ray, and present the semi-empirical response function for each part respectively. Weighted least squares fitting methods is used to fit Gamma spectra of Co-60 and Cs-137 sources measured by CsI(Tl) crystal detector as a typical

## 2 Physical interaction mechanisms and models

There are lots of interactions and scattering following full physics during the detection of Gamma-ray by scintillation detector system. The measured spectra generally contain features that are either highly system dependent, or result from the incident photon and other background rays. Most of these features can be modeled by one kind of suitable semi-empirical response function.

### 2.1 Photoelectron effect

The peak produced by photoelectron effect of Gamma-ray to detect material is usually called full energy peak, as the Gamma-ray energy is absorbed totally. A Gaussian distribution function is adopted to describe the full energy peak.

$$f_1(E_0,E)=\frac{H_1}{\sqrt{2\pi}\sigma_1}\exp(-\frac{(E-E_0)^2}{2\sigma_1^2}) \tag{1}$$

where $E$ is the portion of that energy which is deposited in the detector; $E_0$ is the energy of incident Gamma-ray; $H_1$ is the normalized parameter of pulse height; $\sigma_1$ is the standard deviation of full energy peak.

### 2.2 Pair production

The rest-mass energy ($m_0c^2$) of an Electron is about 0.511 MeV, when the incident Gamma-ray energy exceeds twice (1.02 MeV), the photo will disappear, and Electron-positron pair will be energetically possible produced. In the interaction, the incident photon disappears and is replaced by an electron positron pair. The created positron loosed its kinetic energy and subsequently annihilates with an electron creating two 0.511 MeV Gamma rays that may or may not escape from the detector. So, two functions are adopted to describe the single escape peak and double escape peak respectively.

$$f_2(E_0,E)=\frac{H_2}{\sqrt{2\pi}\sigma_2}\exp(-\frac{(E-(E_0-0.511))^2}{2\sigma_2^2}) \tag{2}$$

$$f_3(E_0,E)=\frac{H_3}{\sqrt{2\pi}\sigma_3}\exp(-\frac{(E-(E_0-1.02))^2}{2\sigma_3^2}) \tag{3}$$

Equation (2) and (3) indicate the single and double escape peak respectively. $H_2$ and $H_3$ are the normalized parameter of pulse height; $\sigma_2$ and $\sigma_3$ are the standard deviation of each peak.

### 2.3 Compton scattering

The incident Gamma-ray interacts with electron in the material; loses its energy and change its direction. The scattered photon energy can be calculated by

$$E' = \frac{E_0}{1+(1-\cos\theta)E_0/m_0c^2} \tag{4}$$

where $\theta$ is the angle between the directions of scattered photon with respect to its original direction. For this part, we take the function proposed by reference [8] as an experience function.

$$f_4(E,\mathrm{E}_0) = H_4\left[\left(\frac{E_0}{E'}\right)+\left(\frac{E'}{E_0}\right)-1+\cos^2\theta\right]erfc\left[\frac{E-A}{\sqrt{2}\sigma_4}\right] \qquad E \le E_c \tag{5-1}$$

$$f_4(E,\mathrm{E}_0) = 0 \qquad E > E_c \tag{5-2}$$

Where $H_4$ and A are fitting parameters; $E' = E_0 - E$ and is equal to the photon energy.

$\cos\theta = 1+\left(\frac{m_0c^2}{E_0}\right)-\left(\frac{m_0c^2}{E'}\right)$ and $E_c = E_0/\left(1+\frac{m_0c^2}{2E_0}\right)$.

### 2.4 Flat continuum

A flat continuum from zero to full energy peak is existed in the spectra, which possibly resulted by electron noise related to incoming Gamma-ray. A kind of function was used in the DRF model for Si(Li),Si(PIN),SDD, HPGe, NaI detectors, and a similar model is observed for CsI(Tl) and NaI(Tl) scintillation detector.

$$f_5(E,E_0) = H_5 erfc\left(\frac{(E-E_0)}{\sqrt{2}\sigma_1}\right) \qquad E \le E_0 \tag{6-1}$$

$$f_5(E,E_0) = 0 \qquad E > E_0 \tag{6-2}$$

where $H_5$ is a normalized parameter. Such a function is obtained by a constant function folded with a Gaussian function.

### 2.5 Exponential tail

There exists an exponential tail on the low energy side of the full energy peak. The representative function is given by

$$f_6(E,E_0) = H_6 \exp\left(\frac{(E-E_0)}{\sqrt{2\pi}\sigma_1\beta}\right) erfc\left(\frac{(E-E_0)}{\sqrt{2}\sigma_1}+\frac{1}{2\beta}\right) \tag{7}$$

where $\beta$ is the slope of this exponential tail. Some studies has adopted two exponential tails to describe this part in different detector, either of them are accepted during the applications.

## 3 Experiment and Spectra fitting

For some applications has reported about NaI(Tl) detector, we choose CsI(Tl) crystal as a typical example here. Eight pieces of CsI(Tl) crystal are employed in the Gamma-ray detector system, and each piece of CsI(Tl) is a cubic part with 24×24×24($mm^3$) in diameter. Gamma-ray emission sources are measured by this system. Before using the DRF model to fit measured Gamma-ray spectra, we should calculate Gaussian broaden parameter$\sigma$ ($\sigma$1 in equation (1)) at first, which is called Gaussian standard deviation.

The parameter $\sigma$ has a similar meaning to FWHM, as they are have a linear relationship, and both can reflect the energy resolution of detector.

$$\sigma = FWHM / \sqrt{8\ln 2} = FWHM / 2.3548 \quad (8)$$

Parameter $\sigma$ can be also calculated by DRF fitting process, but in our former studies, we have presented a kind of new calculation method reported in reference [9] which has made some successful applications. In this study, we have measured the energy range from 0.059MeV (Am-241) to 1.408MeV (Eu-152) Gamma-ray sources to calibrate the standard deviation. While the calibration function has some different formats, and each of them can be accepted as shown in table 1 and figure 1.

**Table 1 Gaussian standard deviation fitting functions and fitting parameters**

| | $\sigma$ | a | b | c | $R^2$ |
|---|---|---|---|---|---|
| Polynomial function | $\sigma = a + bE + cE^2$ | 0.0154242 | 0.0129391 | 0.00681806 | 0.9843 |
| Square root function | $\sigma = a\sqrt{b(E + cE^2)}$ | 0.00995896 | 0.000229984 | 1.97182 | 0.9403 |
| Exponential function | $\sigma = aE^b$ | 0.0367518 | 0.475878 | -- | 0.8885 |

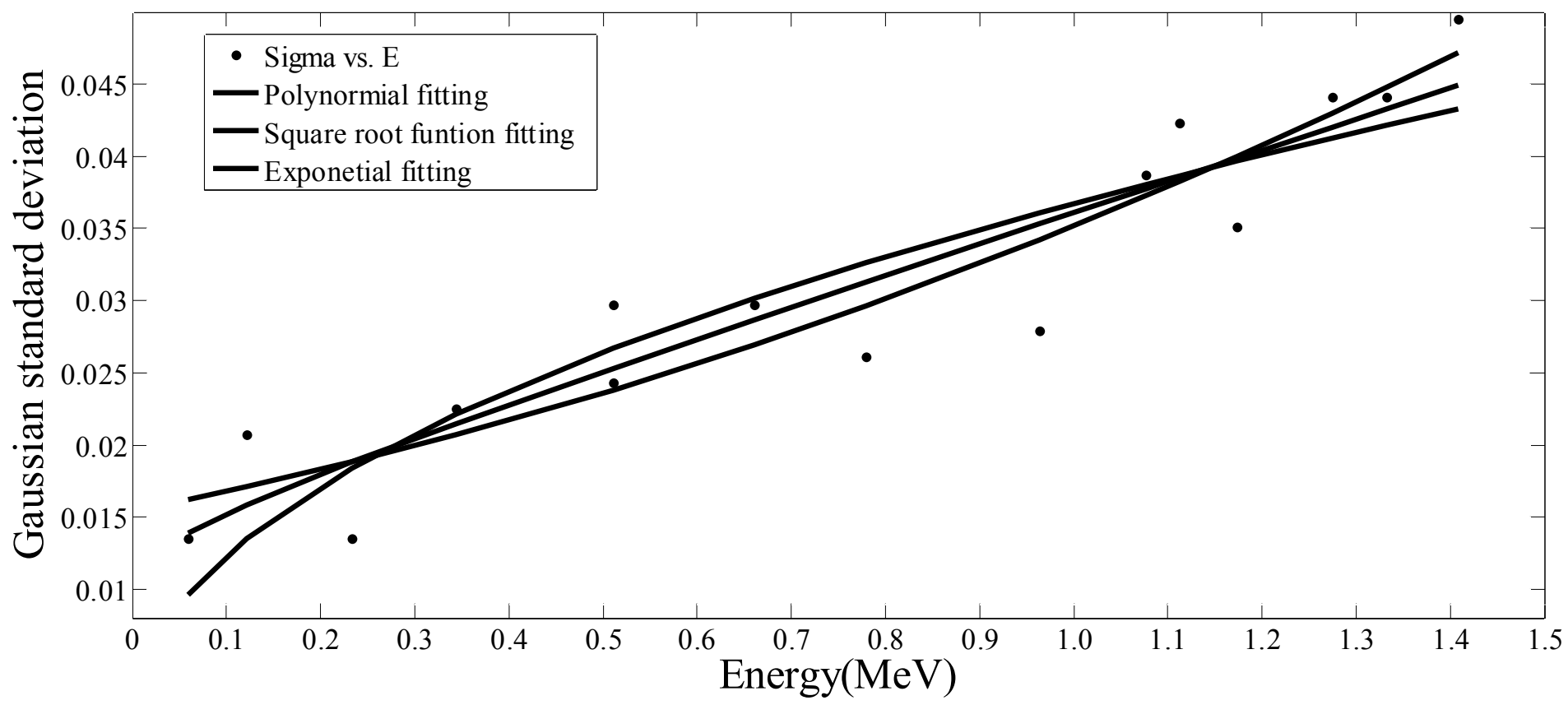

**Figure 1 Gaussian standard deviation fitting results by different functions**

After comparison above, we found that the polynomial fitting function is a good choice. The fitting results of Gaussian standard deviation can be used in DRF model, which can reduce fitting parameters and also can get a much better fitting curve. Two kinds of common Gamma-ray sources: Co-60 and Cs-137 were measured by CsI(Tl) detector, and their spectra are fitted by the DRF model R($E_0$,$E$), Gaussian standard deviation calculated above is also applied simultaneously.

$$R(E_0, E) = \sum_{i=1}^{6} f_i(E_0, E) \quad (9)$$

Though the entire detector response function can be formulated as a probability distribution function (PDF), the whole DRF function can be written as a sum of each part, because multichannel analyzer spectra is a discrete energy form which is more appropriate to account for the accumulated counts in each channel[11].

Weighted least squares (WLS) method is used during the fitting procedure; and this is different from linear least squares fitting method. WLS fitting method can consider the own weight for each channel, which can get more reliable results.

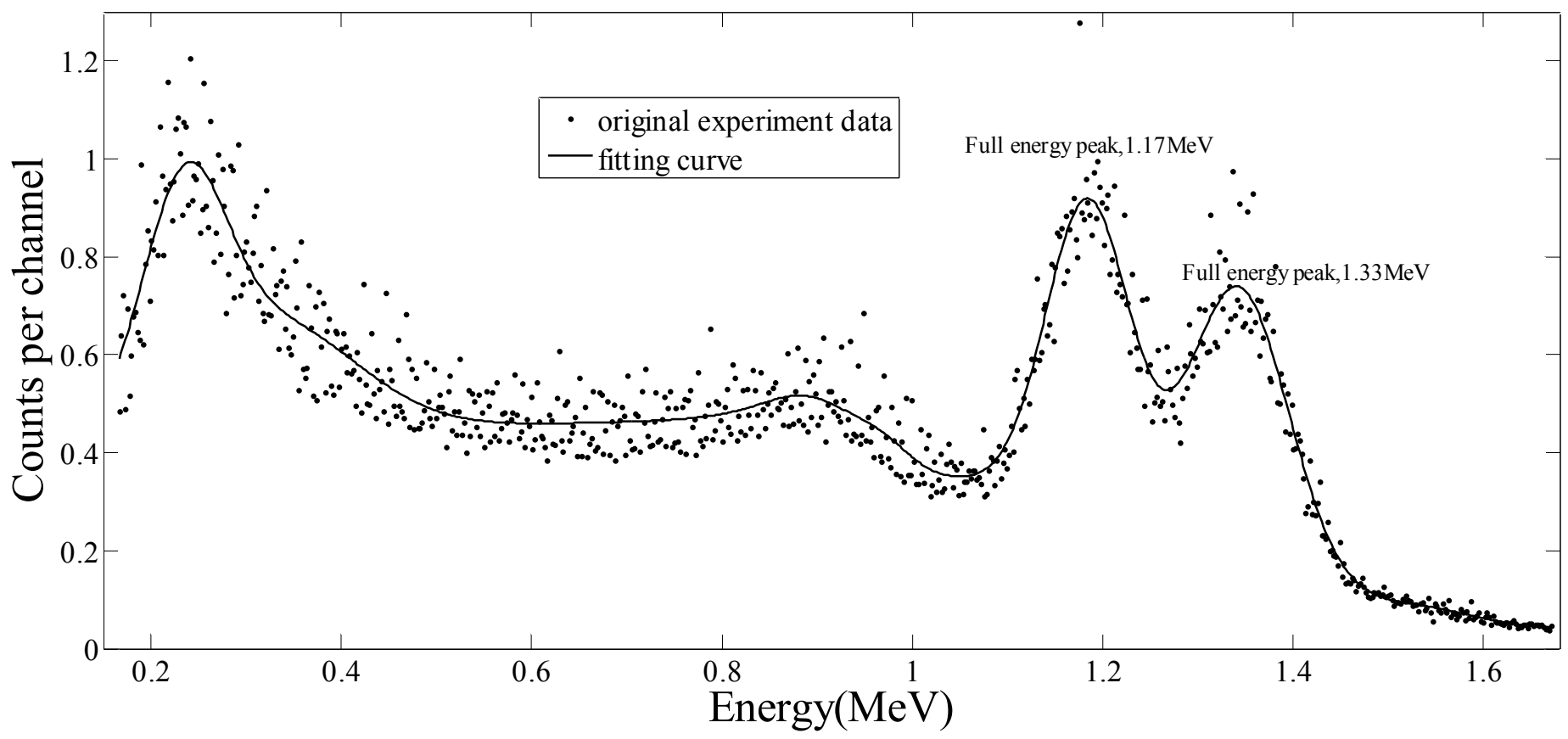

**Figure 2 Fitting results of Gamma-ray spectra of Co-60 source**

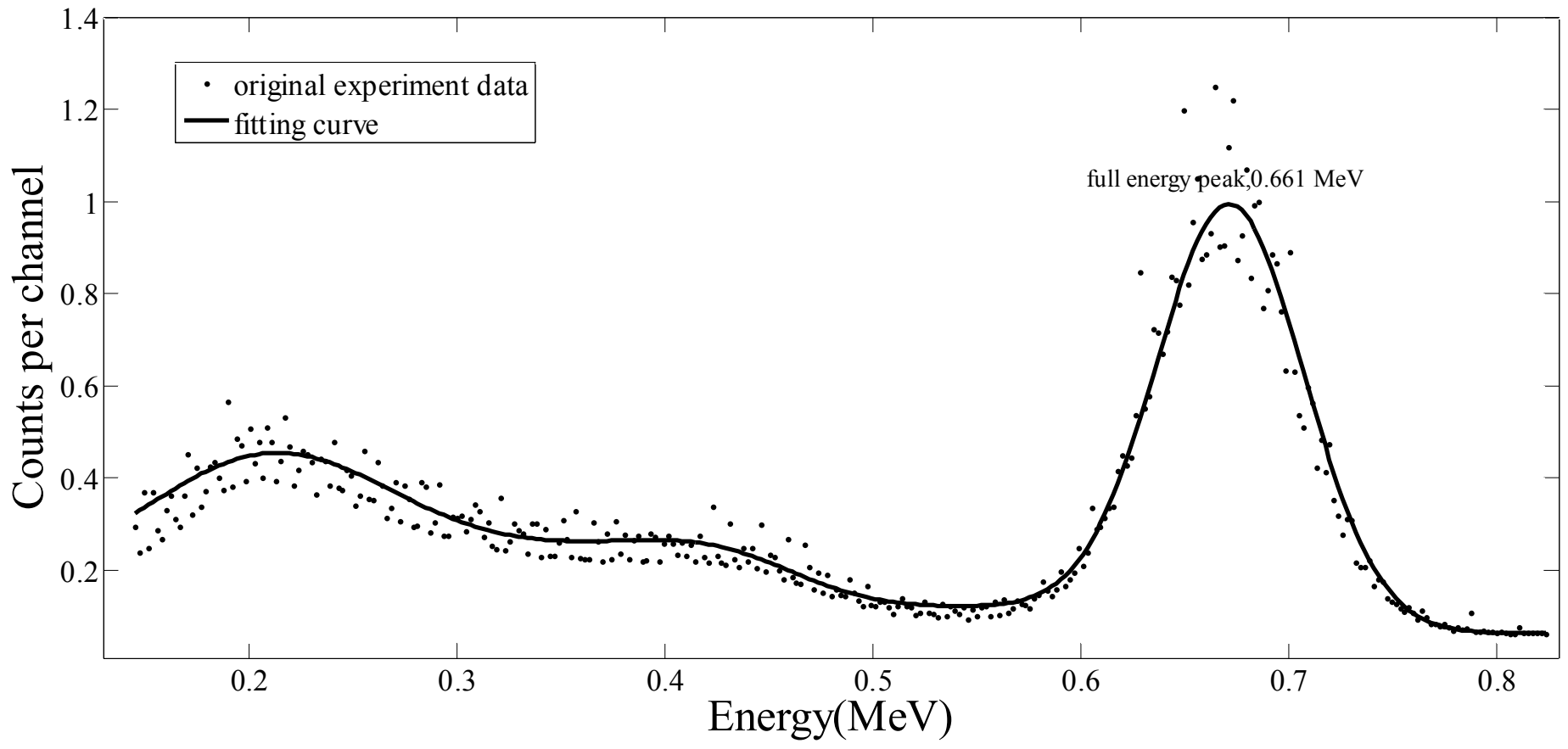

**Figure 3 Fitting results of Gamma-ray spectra of Cs-137 source**

The comparisons between experiments and DRF fitting results are plotted in Figs.2-3. All spectra are normalized to their smoothed highest peaks. Reduced chi-square ( $\chi_r^2$ ) values obtained for each source are 0.63 and 0.83 respectively, so the fitting are generally excellent for these two sources. The peaks around 0.15–0.3 MeV are backscattering peaks, which originate from the Gamma-rays that scattered back from surrounding materials. The back scattering peak is not a legit part of DRF, and we just use a simple Gaussian and tail function to fit it. As the whole DRF model contains four or six parts of each full energy peak, so we can calculate the net area (integral of $f_1$) of full energy peak very easily after the fitting. The fitting results can also show the contribution for each of the features as discussed in sect.2, which can help people understand the complex Gamma-ray spectra from CsI(Tl) detector much deeper. Results in figure 2 and figure 3 are reflected that the Gaussian standard deviation of full energy peak is reasonable.

## 4. Discussion and summary

In this paper, a semi-empirical response function of scintillation detector is presented for Gamma-rays. As analyzed in Sect.2, there are six parts in total for high energy Gamma-rays, which is much more simply and reasonably as compared to other Ge-series detector or Si-series detector. The DRF model contains five different fundamental features; each is described by one or two mathematical expressions that extend over its entire range. There are ten parameters in total, and their values do not require any prior knowledge or assumptions regarding the spectra shape. If people want to reduce the number, they can calculate the Gaussian standard deviation separately as shown in Sect.3. The mathematical form of DRF model can be generally applied to any large volume or array CsI(Tl) detectors. Our primary use of the response function of CsI(Tl) detector is in radio-isotope identification technology, such DRF model can help us understand emission Gamma-rays more accurately. But not only that, the method to establish DRF model for different detectors is very important to take in-deeper and accurate analysis for Gamma or X-rays.